# A Deterministic Binary Fingerprinting Framework with Zero-Trained Feature Extraction for Sparse Count Matrices

*MMTB: multi-threshold thermometer fingerprints and Hamming geometry*

*A Preprint*

Lei Zhao[1], Fujin Huang[2], Ling Kang[1], Quan Guo[*1]

[1] Neusoft Research Institution, Dalian Neusoft University of Information, Dalian, China

[2] The Fourth Research Laboratory, No. 760 Research Institute, China State Shipbuilding Corporation Limited (CSSC), Dalian, China

[*] Corresponding author's email address: guoquan@neusoft.edu.cn

# Abstract

Sparse count matrices from single-cell transcriptomes to k-mer profiles and document-term frequencies are conventionally analyzed via PCA-reduced graph clustering or iterative optimization in continuous embedding spaces. We introduce MMTB, a deterministic non-learned binary representation framework that requires no label supervision, model fitting, or gradient-based optimization. Column-wise Min-Max normalization followed by fixed cutoffs maps each sample to a thermometer fingerprint whose Hamming distances show empirical correspondence with normalized L1 distances, with Pearson correlation approximately 0.92 on single-cell RNA-seq pairs.

In a favorable three-cell-line mixture, the 3-threshold fingerprint achieves NMI of 0.99 at 188 bytes per cell. Under a fair Hamming nearest-neighbor graph plus Leiden readout, MMTB approaches PCA plus Leiden on this coarse task. On challenging tissue-like annotations, continuous pipelines often lead; PBMC Seurat NMI is 0.39 for MMTB versus 0.49 for Scanpy, underscoring that MMTB is suited for coarse-grained separation and resource-constrained deployments rather than fine-grained subtype discovery or as a general replacement for continuous embeddings.

Relative to dense float32 representations, MMTB fingerprints reduce memory by approximately 10-fold while providing fixed-width Hamming-indexable codes. PCA30 embeddings and sparse CSR may be smaller; we do not claim universal compression. A label-free suitability score is provided as a deployment guideline, not a performance predictor.

**Keywords:** binary representation, sparse count matrix, thermometer encoding, Hamming distance, zero-training (non-learned) feature extraction, single-cell genomics

# 1. Introduction

High-dimensional sparse count data are almost always represented in continuous space before analysis: PCA or graph embeddings for single-cell RNA sequencing [1,5,6], dense float features for text and k-mer profiles. Learned binary hashes (ITQ [8], SDH [9], SimHash [20]) target continuous embeddings and require training; they do not define a portable representation for non-negative, zero-inflated count matrices. As a training-free binary baseline, we include SimHash (random hyperplane projections on log1p HVGs with Hamming agglomerative readout) in Supplementary Note 6 (Table S20).

Zero-training here refers narrowly to the fingerprint construction stage. MMTB uses no label supervision, gradient-based optimization, or learned codebook. Column-wise Min-Max requires a single scan of the data to obtain per-feature minima and maxima, a lightweight data dependency that is not a trained transformation.

We propose MMTB as a resource-efficient optional binary feature layer for sparse count matrices: a deterministic encoding with non-learned feature extraction for settings where compact storage, Hamming-indexable geometry, and training-free representation construction matter. Column-wise Min-Max scaling and nested thermometer cutoffs produce a fixed-length binary code whose Hamming metric tracks ordinal within-feature structure under suitable conditions (Section 2.10). Downstream Ward linkage, Hamming kNN, or Hamming-graph Leiden read out global or local structure; MMTB itself is a feature extractor, not an end-to-end clustering pipeline. On a favorable low-K mixture (CellBench), Ward-on-bits can match or exceed graph baselines at 188 B/cell; on tissue-like annotations, continuous pipelines often remain preferable. Compact fingerprints yield approximately 10-fold reduction relative to dense 500-dimensional float32 HVG features (2000 B/cell). PCA30 embeddings (120 B/cell) are a different continuous representation and can be smaller than packed 3-T (188 B/cell). Sparse CSR storage of the original matrix may be smaller still depending on nnz; we do not claim compression versus CSR. A label-free suitability score S is a conservative "worth testing" checklist (Section 3.1), not a predictor used as primary evidence. MMTB is not a replacement for PCA or graph-based biological clustering. It is a deterministic, compact, non-learned binary representation layer that enables Hamming-indexable analysis of sparse count matrices under resource constraints.

We benchmark 12 sparse count domains (Table 4). Five are flagged as highly suitable ($S \geq 80$, $\bar{E} \geq 0.20$); the remaining seven are boundary tests and negative controls. MMTB is a complementary representation strategy for scenarios where deterministic inference, compact storage, and training-free feature extraction are priorities. It is not intended to replace continuous embedding pipelines, for example PCA plus graph clustering, in all settings, and our benchmarking explicitly identifies conditions where it is most versus least suitable.

Contributions. (1) A deterministic non-learned binary representation for sparse counts with fixed thermometer cutoffs and Hamming-indexable codes (Section 1.2). (2) An empirical geometric correspondence account linking thermometer Hamming distance to normalized L1 (Section 2.10), not a metric-preservation theorem. (3) Representation-level evaluation under a shared Leiden readout (Table 8), with candid favorable and limiting author-annotated cases. (4) A practical suitability score S as a deployment guideline only (Section 3.1), not an optimization criterion.

### 1.1 Related Work

Binary codes and thermometer encodings. LSH and SimHash [20,21] approximate similarity in continuous spaces; product quantization [23] uses learned codebooks. Thermometer bits originate in quantized networks [22]. Existing thermometer codes are typically learned per layer; MMTB instead uses global presets on MinMax-normalized count features, which are deterministic, shared across features, and free of codebook fitting (Section 2.9–2.11). As a training-free binary baseline, we include SimHash (random hyperplane projections on log1p HVGs with Hamming agglomerative readout) in Supplementary Note 6 (Table S20).

Distinguishing MMTB from related approaches. Learned embeddings optimize parameters and yield continuous latents; hashing methods approximate similarity via learned or randomized projections; single-cutoff discretization often discards ordinal intensity. MMTB is a complementary representation strategy: deterministic and training-free, without a learned codebook, retaining multi-threshold ordinal information for sparse-count matrices with adequate threshold discrimination.

Single-cell and cross-domain counts. scRNA-seq pipelines often binarize for feature selection but seldom cluster solely in Hamming space [6,7]. Category II provides consistency checks against standardized reference partitions.

### 1.2 Conceptual distinction from conventional discretization and hashing

Binary expression matrices, threshold discretization, and Hamming-space downstream tools are established individually. MMTB's contribution is a fixed, deployment-oriented encoding for non-negative sparse counts. Column-wise Min-Max maps each feature to [0,1]; nested thermometer cutoffs explicitly encode ordinal intensity levels as a short bit stack; and the resulting fixed-length code is consumed under Hamming geometry without learned projections or codebooks. Unlike single-bit median or quantile binarization, nested thresholds retain multi-level intensity structure. Unlike SimHash, ITQ, or product quantization, the map does not rely on random projections or trained quantizers. Adaptive or rank-based binning can be useful locally, but they do not by themselves yield the same portable, fixed-width Hamming-indexable fingerprint under a single non-learned preset. **Table 1** summarizes the distinction.

**Table 1.** Conceptual distinction from conventional discretization and hashing.

| Method | Learned parameters | Requires labels | Explicit ordinal intensity | Fixed-length binary output |
|---|---|---|---|---|
| MMTB | No | No | Yes | Yes |
| Median binary | No | No | No | Yes |
| SimHash | Random projection | No | No | Yes |
| ITQ | Yes | No | Depends | Yes |
| Product quantization | Yes | No | No | Yes |

# 2. Methods

## 2.1 Overview and Problem Formulation

MMTB produces a compact binary representation that can support multiple downstream analyses. In this work, clustering is used to evaluate global structural preservation, whereas Hamming k-NN is used to assess local neighborhood preservation (**Figure 1**). The label-free suitability score (Section 3.1) serves as a practical deployment guideline for choosing between 1-T and 3-T representations before benchmarking.

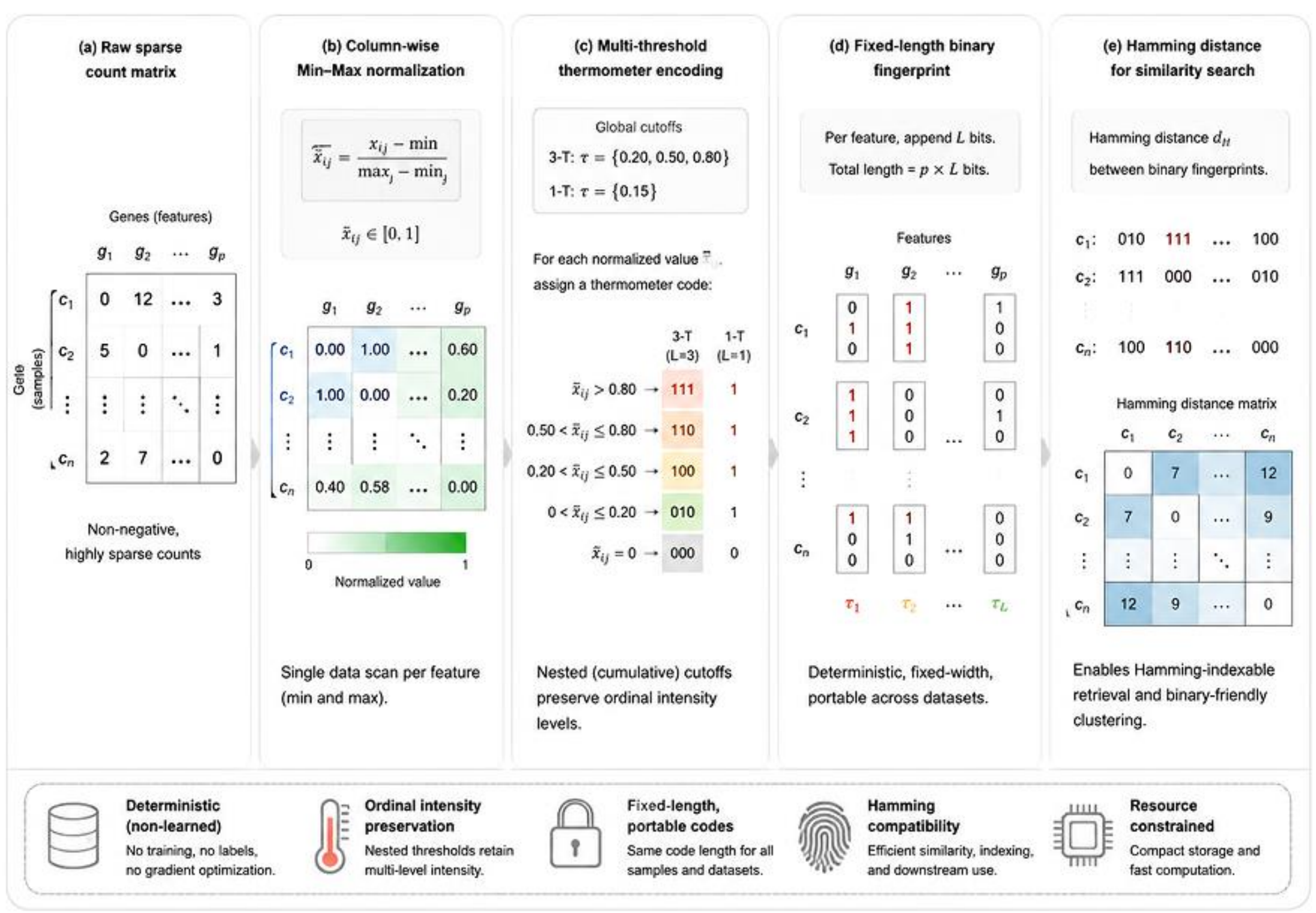

**Figure 1.** Overview of the MMTB encoding pipeline. (a) Raw sparse count matrix with non-negative entries. (b) Column-wise Min-Max normalization maps each feature to [0,1] using a single data scan. (c) Multi-threshold thermometer encoding with global cutoffs (3-T: {0.20, 0.50, 0.80}; 1-T: {0.15}) assigns nested binary codes that preserve ordinal intensity levels. (d) Per-feature codes are concatenated into a fixed-length binary fingerprint of p × L bits. (e) Pairwise Hamming distances between fingerprints enable efficient similarity search, indexing, and clustering. MMTB is deterministic, non-learned, and well-suited for resource-constrained, Hamming-indexable analysis of sparse count data.

Given a data matrix $X \in \mathbb{N}^{N\times d}$ where $N$ is the number of samples (cells) and $d$ is the number of features (e.g., genes, k-mers, or terms), with entries $x_{ij} \in \mathbb{N}_0$ representing UMI counts. Each sample $x_i \in \mathbb{N}_0^d$ is a non-negative integer vector. The goal is to assign each sample to one of $K$ clusters or predict its label via nearest-neighbor retrieval.

MMTB converts each sample $x_i$ into a compact binary fingerprint $b_i \in \{0,1\}^{d\cdot m}$ where $m$ is the number of thresholds (typically $m = 3$). Clustering applies Agglomerative Ward on $b_i$ treated as Euclidean vectors (for $\{0, 1\}$ bits, squared Euclidean distance equals Hamming distance). Local neighborhood preservation uses Hamming distance and kNN majority vote (Section 2.3).

## 2.2 Fingerprint Construction

Each feature $j$ is independently normalized to [0,1] using Min-Max scaling:

$$\hat{x}_{ij} = \frac{x_{ij} - \min_i x_{ij}}{\max_i x_{ij} - \min_i x_{ij} + \epsilon} \tag{1}$$

where $\epsilon = 10^{-8}$ is a small constant to prevent division by zero. This normalization preserves the rank-order structure of each feature while ensuring all values are comparable across features.

Cross-validation (neighborhood readout). In supervised benchmarks (Section 3.5), Min-Max parameters for each feature $j$ are computed from the training fold only: $\min_j$, $\max_j$ use training rows, then Eq. (1) is applied to both training and held-out test rows of that fold. This prevents test-fold expression ranges from leaking into normalization.

For each normalized feature $\hat{x}_{ij}$ and a set of thresholds $\mathcal{T} = \{\tau_1, \tau_2, \dots, \tau_m\}$, MMTB produces $m$ binary bits:

$$b_{ij}^{(t)} = \mathbb{1}[\hat{x}_{ij} \geq \tau_t], t = 1,2,\dots,m \tag{2}$$

The complete fingerprint for sample $i$ is the concatenation of all $m$ bit vectors:

$$b_i = \left[b_{ij}^{(1)} \,|b_{ij}^{(2)} \,| \cdots \,|b_{ij}^{(m)}\right]_{j=1}^{d} \in \{0,1\}^{d \cdot m} \tag{3}$$

For the training-free representation benchmark, we use preset thresholds $\mathcal{T}_3 = \{0.20, 0.50, 0.80\}$ (3-T) and $\mathcal{T}_1 = \{0.15\}$ (1-T), representing low, medium, and high expression levels. No threshold search is performed on any split.

For two binary fingerprints $b_i$ and $b_j$, the normalized Hamming distance is:

$$\text{Hamming}(b_i, b_j) = \frac{1}{d \cdot m} \sum_{l=1}^{d \cdot m} \mathbb{1}[b_{il} \neq b_{jl}] \tag{4}$$

The Hamming distance can be computed efficiently using bitwise XOR operations:

$$\text{Hamming}(b_i, b_j) = \frac{\text{popcount}(b_i \oplus b_j)}{d \cdot m} \tag{5}$$

where $\oplus$ denotes bitwise XOR and popcount counts the number of set bits. This enables highly efficient computation on CPU (using SSE/AVX instructions) and GPU.

## 2.3 Downstream Readouts

Given binary fingerprints $b_i \in \{0,1\}^{dm}$, clustering uses sklearn AgglomerativeClustering with Ward linkage and Euclidean metric on the bit vectors (benchmark default). Ward minimizes the increase in within-cluster variance when merging; for binary features, $|\boldsymbol{\mu}_A - \boldsymbol{\mu}_B|_2^2$ is proportional to the squared Hamming distance between cluster centroids. This is the same linkage as Agg Ward (PCA30) baselines (Section 3.4) but applied in MinMax+binary space rather than PCA-reduced log1p coordinates.

$$\Delta(A, B) = \frac{n_A n_B}{n_A + n_B} |\boldsymbol{\mu}_A - \boldsymbol{\mu}_B|_2^2 \tag{6}$$

where $A$ and $B$ are clusters with sizes $n_A$, $n_B$ and centroids $\boldsymbol{\mu}_A$, $\boldsymbol{\mu}_B$ in fingerprint space. An optional Hamming-distance + average-linkage backend exists in code but is not used in reported benchmarks.

For neighborhood readout, MMTB uses k-nearest neighbors with Hamming distance (lazy learning: no weight training; $k$ is a fixed preset):

$$\hat{y}_i = \mathrm{mode}\big(y_j : j \in \mathcal{N}_k(b_i)\big) \tag{7}$$

where $\mathcal{N}_k(b_i)$ is the set of $k$ nearest neighbors of $b_i$ in the training set under Hamming distance. The mode (majority vote) breaks ties arbitrarily.

## 2.4 Design and Capacity

MMTB applies column-wise Min-Max to log1p HVG features so that fixed thresholds $\tau \in (0,1)$ have portable meaning across features (Eq. 1). log1p before Min-Max stabilizes count dynamic range while preserving ordinal structure within features. Comparisons with z-score, CPM/TPM, and CLR are summarized in Supplementary Note 7.

MMTB 3-T comprises $O(Nd)$ linear Min-Max plus binarization followed by the selected downstream clustering algorithm on binary fingerprints (Supplementary Note 5). PCA30 baselines combine $O(Nd)$ scaling, truncated PCA ($O(N \cdot d \cdot D)$ with $D = 30$), and a chosen clustering algorithm; their cost profile differs from fingerprint generation (Supplementary Note 5). Further details are in Supplementary Note 7.

Thermometer bits count threshold crossings per feature; aggregate Hamming tracks normalized L1 when discrimination is adequate (Supplementary Note 3). Formal lemmas and Idealized Monotonicity are in Supplementary Note 7; Observation 2 states the Hamming-L1 envelope.

Column-wise Min-Max (Eq. 1) maps each feature to [0,1]. Nested thermometer bits encode ordinal levels; per-feature Hamming equals the level gap. With ordered thresholds $\tau_1 < \cdots < \tau_m$, nested bits form a thermometer code whose per-feature Hamming contribution equals the ordinal level gap:

$$\sum_{t=1}^{m} \mathbb{1}[b_{ij}^{(t)} \neq b_{kj}^{(t)}] = |L_j(i) - L_j(k)| \leq m \quad (8)$$

where $L_j(i) = \sum_{t=1}^{m} b_{ij}^{(t)}$ is the ordinal thermometer level for feature $j$ in sample $i$. Summing over features yields H(i,k), a multi-resolution count of threshold crossings. Component-wise dominance rarely holds on real count matrices; the practical link is empirical Hamming-L1 checks (Supplementary Note 3). Empirically, Hamming and normalized L1 correlate strongly on PBMC 4k random pairs: $r = 0.921$, $R^2 = 0.849$, $\rho = 0.915$ (Supplementary Note 3).

Each feature contributes at most m bit flips, yielding the normalized envelope:

$$\text{Hamming}(b_i, b_k) = \frac{H(i,k)}{dm} \leq \frac{1}{d}\sum_{j=1}^{d} \min\left(m, \#\{t: \min(\hat{x}_{ij}, \hat{x}_{kj}) < \tau_t \leq \max(\hat{x}_{ij}, \hat{x}_{kj})\}\right) \quad (9)$$

Observation 2 (Hamming-L1 envelope, qualitative). For any samples $i, k$:

$$H(i,k) \leq \sum_{j=1}^{d} \min\left(m, \#\{t: \min(\hat{x}_{ij}, \hat{x}_{kj}) < \tau_t \leq \max(\hat{x}_{ij}, \hat{x}_{kj})\}\right) \quad (10)$$

Hamming is an aggregate envelope of per-feature ordinal separation (Observation 2), not an L1 isometry. Observation 2 provides a qualitative envelope relating aggregate Hamming distance to normalized L1 under idealized assumptions; the bound can be loose when bits are sparse or unbalanced (low $\bar{E}$). It is not a formal guarantee of metric correspondence or neighborhood-order identity. Pearson correlation (e.g., $r \approx 0.92$ on PBMC 4k pairs) indicates empirical geometric correspondence when discrimination is adequate, not isometry. The practical backbone is multi-cohort empirics (Supplementary Note 3 / Table S21) together with the representation-level evaluations in Section 4.1.

Nested thresholds define a thermometer code with ordinal capacity at most $\log_2(m+1)$ bits per feature:

$$I_{\max}(L_j) = \log_2(m+1) \text{ bits/feature} \quad (11)$$

**Table 2** reports the theoretical capacity and empirical clustering NMI for the two preset configurations. The 3-T preset stores three bits per feature but encodes two bits of ordinal information per feature (four ordinal levels), while 1-T stores one bit per feature with binary capacity. The NMI values correspond to Ward-on-bits clustering on PBMC 4k against the Ward reference partition (Supplementary Note 6); these are illustrative of the capacity-performance trade-off rather than primary biological validation.

**Table 2.** Encoding capacity and downstream clustering NMI for 1-T and 3-T presets on PBMC 4k (Ward reference partition).

| Encoding | Ordinal states | Capacity $I_{\max}$ | Stored bits/feature | Clustering NMI |
|---|---|---|---|---|
| 1-T ($\tau = 0.15$) | 2 | $\log_2 2 = 1$ bit | 1 | 0.4769 |
| 3-T ($\{0.20, 0.50, 0.80\}$) | 4 | $\log_2 4 = 2$ bits | 3 | 0.8349 |

Under a uniform ordinal model,

$$H(L_j) = \log_2(m+1) \tag{12}$$

Single-bit Bernoulli entropy is at most 1 bit:

$$H(B_1) = -p_j \log_2 p_j - (1-p_j)\log_2(1-p_j) \leq 1 \text{ bit} \tag{13}$$

Single-bit entropy collapses when activation rates approach 0 or 1; practice therefore depends on balanced cutoffs. Recoverable cluster information is bounded by $d\log_2(m+1)$:

$$I(L;C) \leq H(L) = \sum_{j=1}^{d} H(L_j \mid L_{<j}) \leq \sum_{j=1}^{d} H(L_j) \leq d\log_2(m+1) \tag{14}$$

Label-free discrimination uses balance scores $s_j(\tau)$ and threshold discrimination $\bar{E}$:

$$s_j(\tau) = \begin{cases} 1 - 2 \mid p_j(\tau) - 0.5 \mid & \text{if } 0.05 < p_j(\tau) < 0.95 \\ 0 & \text{otherwise} \end{cases} \tag{15}$$

$$E(\tau) = \frac{1}{|\mathcal{J}_\tau|} \sum_{j \in \mathcal{J}_\tau} s_j(\tau), \bar{E} = \frac{1}{m} \sum_{t=1}^{m} E(\tau_t) \tag{16}$$

Mean bit entropy $\bar{H}_{\text{bit}}$ complements $\bar{E}$:

$$\bar{H}_{\text{bit}} = \frac{1}{m} \sum_{t=1}^{m} \frac{1}{d} \sum_{j=1}^{d} H(B_t^{(j)}), B_t^{(j)} = \mathbb{1}[\hat{x}_{\cdot j} \geq \tau_t] \tag{17}$$

where $H(B_t^{(j)})$ is the Bernoulli entropy of bit $t$ for feature $j$. For PBMC 3k 3-T, $\bar{E} = 0.263$ and $\bar{H}_{\text{bit}} = 0.083$. $\bar{E} < 0.20$ indicates collapsed Hamming discrimination.

Fixed thresholds partition $[0, 1]$ into at most $(m+1)$ ordinal bins; reconstruction error is bounded by the widest bin:

$$\mid \hat{x}_{ij} - \tilde{x}_{ij} \mid \leq \max_{0 \leq t \leq m} (\tau_{t+1} - \tau_t) \text{ with } \tau_0 = 0, \tau_{m+1} = 1 \tag{18}$$

For 3-T $\{0.20, 0.50, 0.80\}$, the maximum bin width is 0.30. 1-T stores fewer bits at coarser resolution, an explicit capacity-storage trade-off.

Preset thresholds are engineering heuristics selected on the PBMC 4k development panel to maximize label-free $\bar{E}$ (Section 3.1), not to maximize NMI against labels or Ward references. PBMC 4k is therefore a development cohort for threshold choice and is not framed as primary biological validation; Category II consistency for PBMC 4k appears only in Supplementary Note 6. Grid combinations that raise NMI on a labeled or Ward target (e.g., {0.30, 0.50, 0.70} in Supplementary Table S1) were rejected because they amount to label- or reference-guided tuning and are inconsistent with the non-learned feature extraction protocol. The cutoffs avoid extremes (<0.05 or >0.95): 0.20/0.50/0.80 span low/mid/high relative expression with stable mid-range Ē, while 0.15 for 1-T targets a single informative presence bit below the mid-scale 0.20 used in the 3-T ladder.

## 2.5 Notation Summary

Table 3 defines all symbols used in the equations throughout Section 2. References to this table are included parenthetically where each symbol first appears in the main text; readers may consult Table 3 for a consolidated summary.

**Table 3.** Symbols used in Equations (1)–(19).

| Symbol | Meaning |
|---|---|
| $N$ | Number of samples (cells) |
| $d$ | Number of features (e.g., HVG genes, k-mers) |
| $m$ | Number of binarization thresholds ($m = 3$ for 3-T, $m = 1$ for 1-T) |
| $x_{ij}$ | Raw or log1p expression of sample $i$, feature $j$ |
| $\hat{x}_{ij}$ | Min-Max normalized value in $[0,1]$ (Eq. 1) |
| $\epsilon$ | Small constant ($10^{-8}$) preventing zero division |
| $\tau_t$ | The $t$-th binarization threshold (Eq. 2) |
| $b_{ij}^{(t)}$ | Binary bit at threshold $t$ (Eq. 2) |
| $b_i$ | Concatenated fingerprint for sample $i$ (Eq. 3) |
| $L_j(i)$ | Ordinal thermometer level for feature $j$, sample $i$ (Eqs. 8, 11) |
| $B_t, B_t^{(j)}$ | Binary bit at threshold $t$ (feature $j$); Eq. 13, 17 |
| $H(L_j)$, $H(B_1)$ | Entropy of ordinal level / single bit (Eqs. 12–13) |
| $I(L; C)$ | Mutual information between thermometer levels and cluster labels (Eq. 14) |
| $I_{\max}(L_j)$ | Max information capacity $\log_2(m+1)$ bits/feature (Eq. 11) |
| $\bar{H}_{\text{bit}}$ | Mean per-threshold Bernoulli entropy (Eq. 17) |
| $H(i,k)$ | Hamming bit-disagreement count (Eqs. 8–10) |
| $p_j(\tau)$ | Fraction of cells with $\hat{x}_{\cdot j} \geq \tau$ (Eq. 15) |
| $E(\tau), \bar{E}$ | Per-threshold and mean threshold discrimination (Eq. 16) |
| cpc | Cells per cluster ($N/K$) |
| $K$ | Number of clusters or CV folds |
| $\hat{y}_i$ | Predicted class label (Eq. 7) |
| $\mathcal{N}_k(b_i)$ | $k$ nearest neighbors under Hamming distance |
| NMI, ARI | Normalized Mutual Information[18] and Adjusted Rand Index[19] |

# 3. Implementation and Benchmark

## 3.1 Practical deployment guideline: label-free suitability screening

We target sparse count-like matrices (UMI counts, BoW term frequencies, k-mer rates, species count tables). The unified benchmark spans 12 datasets: 7 scRNA-seq, microbiome 6-mer (MicroPheno), synthetic k-mer, pooled reef-fish ecology counts (Dryad), and NLP negative controls (Sections 4.6, 4.7). Large-scale PBMC 68k (10x donor A, $S = 98$) is evaluated separately in Section 4.9.3. Suitability v2.6 (suitability.py) scores each matrix before benchmarking.

Before benchmarking, each candidate matrix is scored by suitability v2.6 on the same log1p HVG matrix used in evaluation. The score $S \in [0,100]$ sums weighted checks: dimensionality ($d \geq 50$), discrete/count nature (integer ratio), sparsity band, dynamic range, threshold discrimination $\bar{E}$ (Eq. 16), sample size, and cells-per-cluster budget. Two multipliers may reduce the score: a sparse × discrimination gate ($\times\, g$, Eq. 19) when sparsity < 10% and $\bar{E} < 0.25$, and a fine-cluster penalty when $K > 10$. This score is an engineering heuristic rather than an optimal statistical estimator; the weighting scheme was designed for conservative deployment rather than theoretical optimality, and $S$ should therefore be interpreted as an applicability screening tool.

$$g = \begin{cases} 1 & \text{if sparsity} \geq 0.10 \text{ or } \bar{E} \geq 0.25 \\ \text{clip}(\bar{E}/0.25, 0.35, 1) & \text{otherwise} \end{cases} \tag{19}$$

**Recommendation gates.** A dataset is Recommended for MMTB when it passes suitability gates: $S \geq 65$, $\bar{E} \geq 0.20$, $d \geq 50$, integer ratio $\geq 0.50$, and (if sparsity < 5%) additionally $\bar{E} \geq 0.25$. Marker-panel exceptions are documented in suitability.py. Recommended indicates expected non-degenerate thermometer codes; it is independent of Experiment Included (**Table 4**). The suitability score is not used for model optimization or performance ranking; it serves only as a deployment guideline.

**Two-tier gates.** Recommended ($S \geq 65$, $\bar{E} \geq 0.20$; Table 4) marks matrices worth trying MMTB. Highly Suitable ($S \geq 80$) is a stricter deployment-screening band for suitability-conditioned analyses in Section 4 (including PBMC 68k where applicable); it is not a substitute for Category I biological validation. Favorable CellBench results are illustrative; MicroPheno and Seurat markers remain limitation exemplars.

**Worth-testing gates versus experiment inclusion. Table 4** separates two decisions: Worth testing / gate label "Recommended" (suitability gates: $S \geq 65$, $\bar{E} \geq 0.20$; an eligibility screen, not a performance guarantee) and Experiment Included (whether the dataset appears in Section 4). PBMC 5k is Worth testing = No ($\bar{E} \approx 0.12 < 0.20, S = 32$) but Experiment Included = Yes; it stress-tests MMTB on a low-discrimination matrix.

**Proposition 1 (Eligibility and bit degeneracy).** If $\bar{E} < 0.20$, then for most thresholds the balance score $s_j(\tau)$ is near zero for most features (Eq. 15). Consequently, the mutual-information upper bound $I(L; C)$ (Eq. 14) is near zero: most bits are nearly constant across cells, and Hamming distance collapses (empirically: Newsgroups $\bar{E} = 0.04$, MMTB NMI $= 0.05$).

**Proposition 2 (Score as label-free screening criterion).** Score $S$ is an engineering heuristic that flags candidate matrices worth testing with MMTB. It is a conservative user-guide checklist; not a formal predictor and not used as primary performance evidence. The suitability score is not used for model optimization or performance ranking; it serves only as a deployment guideline. Paul 2015 can pass Recommended gates yet underperform when $K$ is large. Screening tables remain in Supplementary materials as a user checklist. Future work should validate $S$ on independent collections beyond the present screen set.

**Clarification.** Score $S$ is an engineering heuristic that flags candidate matrices worth testing with MMTB. It is a conservative screening tool for deployment decisions rather than a proven statistical predictor.

**Design principles (why these weights and gates).** The scoring function implements three explicit principles aligned with Section 2.4:

Principle 1 — Threshold discrimination dominates. The largest point block ($S_{\bar{E}} \leq 28$) rewards $\bar{E}$ because Bernoulli entropy (Eq. 13) collapses when $p_j(\tau) \to 0$ or 1: without balanced bits, Hamming distance cannot separate samples. This is why $\bar{E} \geq 0.20$ is a hard eligibility gate; below it, thermometer codes are structurally degenerate (Newsgroups, $\bar{E} = 0.04$).

Principle 2 — Count-like structure and scale. Integer-ratio ($S_{\text{disc}} \leq 15$), sparsity band, dynamic range, and dimensionality ($d \geq 50$) ensure the input resembles the sparse count matrices MMTB was designed for, not arbitrary continuous embeddings. Marker panels and dense k-mer profiles receive profile-specific exceptions (Moignard, MicroPheno) documented in suitability.py.

Principle 3 — Conservative deployment tiers. Verdict bands follow a fail-safe ladder: $S \geq 80$ (Highly Suitable) requires strong $\bar{E}$ and sufficient sample/cluster budget; stricter than eligibility ($S \geq 65$), which only indicates that testing is warranted. The sparse × discrimination gate $g \in (0,1]$ (Eq. 19) down-weights ultra-sparse, low-$\bar{E}$ matrices before the fine-cluster factor penalizes large $K$. These are heuristic guardrails associated with safer deployment, not optimality proofs.

**Table 4.** Dataset suitability screening

| Dataset | Score | $\bar{E}$ | $K$ | cpc | Recommended | Experiment Included |
|---|---|---|---|---|---|---|
| PBMC 3k (10x Genomics) | 100 | 0.347 | 5 | 529 | Yes | Yes |
| PBMC 4k (10x Genomics) | 105 | 0.393 | 5 | 771 | Yes | Yes |
| PBMC 5k (10x Genomics) | 32 | 0.124 | 5 | 1063 | No | Yes |
| Moignard 2015 (blood differentiation) | 98 | 0.695 | 5 | 787 | Yes | Yes |
| CellBench 10x (3-line mixture) | 80 | 0.298 | 3 | 301 | Yes | Yes |
| CellBench Drop-seq (3-line mixture) | 79 | 0.312 | 3 | 75 | No | Yes |
| Paul 2015 (mouse bone marrow) | 69 | 0.377 | 19 | 144 | No | Yes |
| 20 Newsgroups (BoW, n=2000) | 60 | 0.219 | 20 | 100 | No | Yes |
| RCV1 Reuters (BoW, n=2000) | 16 | 0.059 | 20 | 94 | No | Yes |
| Synthetic k-mer counts (n=2000) | 78 | 0.211 | 5 | 400 | No | Yes |
| MicroPheno HMP body-sites (6-mer, n=1192) | 80 | 0.267 | 5 | 238 | Yes | Yes |
| Dryad pooled reef fish counts (120 sites, 5 regions) | 51 | 0.169 | 5 | 24 | No | Yes |
| PBMC 68k full (10x donor A) | 98 | 0.308 | 11 | 6206 | Yes | Yes |

* Recommendation is a heuristic gate for conservative deployment, not a guarantee of performance. See Section 3.1 for limitations.

**Reference-partition panels.** PBMC 4k (score 105, $\bar{E} \approx 0.39$, cpc $\approx 771$) is the largest suitable Ward-reference panel for ablation and sensitivity (Section 4.3). CellBench 10x ($\bar{E} \approx 0.30$, author annotations) anchors Category I biological validation.

**Why Newsgroups fails.** Although 20 Newsgroups has integer BoW counts and 500 features (base score 59 before gates), $\bar{E} = 0.04$: at every cutoff $\tau \in \{0.20, 0.50, 0.80\}$, nearly all terms are either always absent or always present, so Hamming distance collapses. The sparse × discrimination gate ($\times\ 0.35$) and fine-cluster penalty ($K = 20$) reduce the final score to 17. This matches the theoretical prediction in Section 2.4 and validates screening before applying MMTB to non-expression domains.

## 3.2 Datasets

**Table 5** lists 6 Recommended benchmark cohorts ($S \geq 65$ eligibility gates; six also meet Highly Suitable $S \geq 80$). Additional stress-test and negative-validation datasets appear in Sections 4.6, 4.7.

**Table 5.** Dataset statistics

| Dataset | Cells | Features | Clusters | Suitability | Domain | Source |
|---|---|---|---|---|---|---|
| PBMC 4k (10x Genomics) | 3,857 | 500 | 5 | 105 | scRNA-seq | 10x Genomics |
| PBMC 3k (10x Genomics) | 2,643 | 500 | 5 | 100 | scRNA-seq | 10x Genomics |

| Moignard 2015 (blood differentiation) | 3,934 | 42 | 5 | 98 | scRNA-seq | Moignard et al., Nature Biotechnology 2015 (data/moignard15/) |
|---|---|---|---|---|---|---|
| CellBench 10x (3-line mixture) | 902 | 500 | 3 | 80 | scRNA-seq | Tian et al., Genome Biology 2019 (author cell-line labels) |
| MicroPheno HMP body-sites (6-mer, n=1192) | 1,192 | 4096 | 5 | 80 | microbiome-kmer | Asgari et al., Bioinformatics 2018 (HMP 16S k-mer rates) |
| PBMC 68k full (10x donor A) | 68,265 | 500 | 11 | 98 | scRNA-seq | 10x Genomics fresh donor A (Zheng et al.) |

Evaluation regimes. Datasets with author or experimental annotations are evaluated against those labels; datasets without accepted annotations are evaluated against Ward-derived reference partitions (Section 3.3).

## 3.3 Evaluation reference partitions

**Evaluation reference categories**

Category I — Independent biological labels. CellBench (author cell-line labels), MicroPheno (body-site labels), synthetic k-mer mixtures, and NLP topic labels. These annotations are independent of MMTB and all clustering baselines.

Category II — Ward-derived reference partitions (consistency check). PBMC 3k, 4k, 5k, Paul 2015, and Moignard 2015 lack accepted cell-type annotations. Category II quantifies agreement with standardized reference partitions under a common linkage objective; an algorithmic consistency check, not biological validation.

Category III — Independent secondary annotations. PBMC 3k only: Seurat marker-based cell-type annotation ($K = 9$, Table 6), used solely for secondary validation.

Two evaluation regimes. Datasets therefore fall into two evaluation regimes. Datasets with experimentally or author-provided annotations (e.g., CellBench, MicroPheno) evaluate biological agreement (Category I), whereas datasets without accepted annotations (e.g., PBMC, Paul, Moignard) undergo consistency checks against a standardized reference partition constructed by Agglomerative Ward on the log1p HVG representation. These reference partitions are standardized clustering outcomes for algorithm comparison rather than biological ground truth.

Ward linkage and reference bias. Ward reference partitions on PBMC panels share linkage family with MMTB + Ward but not feature space. We mitigate circularity by: (1) excluding Agg Ward (Log1p) generators from fair ranking (NMI = 1.0 by definition); (2) leading Results with author-annotation cohorts (CellBench, MicroPheno; Section 4.1); (3) re-scoring PBMC 3k against Seurat marker annotation (nine immune classes [6]; Table 6, Category III); (4) including a Scanpy/Seurat-like baseline alongside Leiden (PCA30). Category II remains a consistency diagnostic, not a substitute for Category I.

**Table 6.** PBMC 3k secondary validation: clustering NMI vs Seurat marker annotation (K=9).

| Method | NMI | ARI | Notes |
|---|---|---|---|
| Scanpy/Seurat-like (Leiden) | 0.4887 | 0.3958 | Scale+PCA+Leiden; best on author labels |
| Agg Ward (MinMax 500D) | 0.4742 | 0.3773 | Continuous MinMax space |
| Leiden (PCA30) | 0.4676 | 0.3717 | Graph baseline |
| MMTB 3-T zero-train | 0.3922 | 0.2506 | Binary representation + Ward |
| MMTB 1-T zero-train | 0.2818 | 0.2009 | Single-threshold code |

Category III secondary check. **Table 6** re-scores PBMC 3k against Seurat marker annotation—independent of the Ward reference partition used elsewhere on PBMC panels.

## 3.4 Compared Algorithms

Naive single-bit baseline. For fair comparison in MinMax space, we define Median-bit codes by per-feature median thresholding of MinMax-normalized values (bit $j = 1$ iff $x_{ij} \geq \text{median}_j$), followed by the same Ward-on-bits readout used for MMTB in the Category I comparison table. This is distinct from the fixed 1-T preset $\{0.15\}$.

We compare MMTB 1-T and 3-T non-learned feature-extraction variants against seven baselines implemented in scikit-learn [2] and Scanpy [7]. **Table 7** summarizes the compared algorithms, their storage costs, training requirements, and references.

**Table 7.** Compared algorithms.

| Algorithm | Type | Storage/cell | Training? | Reference |
|---|---|---|---|---|
| MMTB 3-T zero-train | Binary fingerprint + Ward (clf: Hamming kNN) | 188 B | No | This work |
| MMTB 1-T zero-train | Binary fingerprint + Ward (clf: Hamming kNN) | 500 B | No | This work |
| Leiden (PCA30) | Graph community | 120 B | PCA + graph | Traag et al.[5] |
| Scanpy/Seurat-like | Scale+PCA+Leiden | 120 B | PCA + graph | Seurat workflow[6] |
| Spectral (PCA30) | Spectral | 120 B | PCA + eigend. | Ng et al.[3] |
| K-Means (PCA30 / Log1p) | Partition | 120–2000 B | PCA or raw fit | Lloyd[2] |
| Agg Ward (PCA30 / Log1p) | Hierarchical | 120–2000 B | Fitting | Ward[17] |
| RF (Log1p 500D) | Ensemble trees | 2000 B | Yes | Breiman[14] |
| SVM RBF / LR (PCA30) | Kernel / linear | 120 B | Yes | Cortes & Vapnik[16] |
| PCA30 + kNN / Log1p + kNN | Instance-based | 120–2000 B | Yes / No | Cover & Hart[15] |

* Storage reported for the feature representation only. PCA30 stores 30 float32 values (120 B). MMTB 3-T stores 500×3 packed bits (188 B). The relevant baseline for MMTB's compression efficiency is the raw log1p HVG matrix (500×4 = 2000 B), not PCA30.

## 3.5 Experimental Protocol

Reproducibility. All stochastic operations use fixed random_state=42 for train/test splits, PCA, Leiden resolution search, k-means n_init, and subsample indices. Hamming-L1 pair sampling uses the same seed.

Local neighborhood preservation. Hamming kNN on the MMTB binary fingerprint is the natural downstream readout for local structure, reported alongside conventional trained methods (SVM, LR, RF, PCA30+kNN) and training-free baselines (Log1p+kNN). All use the same stratified 5-fold cross-validation (StratifiedKFold, random_state=42, shuffle). Within each fold: column Min-Max (Eq. 1), StandardScaler, and PCA are fit on the training partition only and applied to train and test. MMTB uses fixed preset thresholds and Hamming kNN with fixed $k$; no gradient-based optimization or label-guided threshold tuning. We report mean $\pm$ SD accuracy across the five folds (Tables 8–9). Each cell is evaluated exactly once as test data.

Clustering. All cells are used on the full preprocessed matrix. Clustering assignments are compared to author annotations or Ward reference partitions via NMI and ARI [18, 19]. MMTB uses fixed preset thresholds and Ward linkage on Hamming fingerprints; deterministic given the input matrix. All clustering algorithms (Leiden, Louvain, PhenoGraph, Birch, K-Means, HDBSCAN, Ward, etc.) are downstream consumers of their respective feature representations; NMI scores evaluate representational adequacy, not superiority of any single clustering objective. Section 4.2 reports ablation and sensitivity on PBMC 4k; Section 4.4 repeats clustering on PBMC 4k with 10 stratified 85% subsamples (random_state=42 per repeat) with bootstrap 95% confidence intervals (Table 28).

Leakage prevention. Neighborhood-preservation folds never use test-fold rows to fit Min-Max, Scaler, PCA, or classifiers. MMTB thresholds are fixed constants and are never tuned on hold-out labels. Clustering is unsupervised on the full matrix; only neighborhood-preservation experiments isolate train/test per fold.

Preprocessing. Standard QC (min_genes=200, min_cells=3, mt < 5%), log1p normalization, top 500 HVG selection. All methods receive the same preprocessed data.

# 4. Results

We report an illustrative favorable Category I case (CellBench), then balanced limitation exemplars (MicroPheno body sites; PBMC 3k Seurat markers), binary baselines under a shared Ward-on-bits readout, and a fair graph-style Leiden readout on Hamming-kNN graphs of the binary codes. Category II Ward-reference consistency checks remain in Supplementary Note 6 only.

## 4.1 Representation-level evaluation under a shared graph readout (favorable and limiting author-annotated cases)

We evaluate MMTB primarily as a representation. Subsection 4.1.1 reports Hamming-kNN + Leiden on the binary codes under the same community-detection family used for PCA baselines (**Table 8**); this is the fairest main-text comparison of representation quality. Subsequent subsections place favorable coarse-grained author-label results (CellBench; **Table 9** row) and candid limitation exemplars (MicroPheno; PBMC Seurat) around that fair readout, then binary hashing baselines and neighborhood preservation. MMTB is not a replacement for PCA or graph-based biological clustering. It is a deterministic, compact, non-learned binary representation layer that enables Hamming-indexable analysis of sparse count matrices under resource constraints.

### 4.1.1 Fair representation evaluation: Hamming-kNN + Leiden on binary codes

To separate representation quality from downstream algorithm choice, **Table 8** reports a fair graph-style readout: instead of Ward linkage on binary codes (used in **Table 9**), we build a Hamming kNN graph (n_neighbors=15, metric=hamming) from the MMTB fingerprints and run Leiden community detection. This brings MMTB's feature space under the same graph-clustering paradigm used for PCA baselines, isolating the effect of the binary representation from the effect of the clustering algorithm family.

Comparing MMTB+Ward to PCA+Leiden conflates representation and algorithm family. Table 8 reports a fairer graph-style readout: build a Hamming kNN graph on the binary codes (n_neighbors=15, metric=hamming) and run Leiden. On CellBench 10x, MMTB 3-T Hamming-Leiden NMI = 0.886 (vs PCA+Leiden 0.825; Ward-on-bits 0.988). On PBMC 3k Ward-ref, MMTB Hamming-Leiden NMI = 0.765 nearly matches PCA+Leiden 0.771. On Seurat markers (limitation), MMTB Hamming-Leiden 0.448 still trails PCA+Leiden 0.468. MMTB remains a feature layer; community detection is an optional downstream consumer.

**Table 8.** Fair graph readout — Hamming-kNN + Leiden on binary codes vs PCA30+Leiden / Ward-on-bits (NMI).

| Dataset | Tier | MMTB 3-T | SimHash-500 | Median-bit | Leiden | B/cell |
|---|---|---|---|---|---|---|

| CellBench 10x (3-line mixture) | Category I | 0.9879 | 0.9811 | 0.9797 | 0.8252 | 188\|63\|63\|120 |
|---|---|---|---|---|---|---|
| CellBench Drop-seq (3-line mixture) | Category I | 0.8290 | 0.8531 | 0.8521 | 0.8372 | 188\|63\|63\|120 |
| MicroPheno HMP body-sites (6-mer, n=1192) | Category I | 0.6409 | 0.5781 | 0.7042 | 0.6731 | 188\|63\|63\|120 |
| PBMC 3k (10x Genomics) | Independent PBMC (Ward-ref) | 0.8349 | 0.6874 | 0.3760 | 0.7709 | 188\|63\|63\|120 |

Table columns: MMTB 3-T = thermometer Ward-on-bits; SimHash-500 = random hyperplane binary codes + Ward-on-bits; Median-bit = per-feature median binarization + Ward-on-bits; Leiden = PCA30 + Leiden. B/cell entries correspond to MMTB 3-T | SimHash | Median-bit | PCA30.

#### 4.1.2 Favorable coarse-grained separation: CellBench cell-line labels

**Table 9** summarizes MMTB 3-T vs Leiden on author-annotation cohorts (Category I). Full algorithm rankings appear in Supplementary Note 6.

**Table 9.** Category I summary: author annotations (MMTB 3-T vs Leiden PCA30)

| Dataset | Evaluation target | MMTB 3-T NMI | Leiden NMI | ΔNMI | Storage |
|---|---|---|---|---|---|
| CellBench 10x | Author cell_line ($K = 3$) | 0.9879 | 0.8252 | +0.1627 | 188 B |
| MicroPheno HMP | Author body-site ($K = 5$) | 0.6409 | 0.6731 | −0.0322 | 188 B |

CellBench. Three lung cancer cell lines on 10x and Drop-seq; author annotations independent of any clustering algorithm. This low-K mixture is an easier favorable setting than tissue subtypes. MMTB 3-T Ward-on-bits clustering NMI = 0.9879 on 10x (ΔNMI = +0.1627 versus Leiden). **Figure 2** visualizes the NMI comparison for CellBench 10x and Drop-seq.

CellBench 10x (3-line mixture): MMTB 3-T local-neighborhood readout (Hamming kNN) 99.67% ± 0.44% (5-fold CV); best MMTB clustering NMI 0.9879 (188 B). Leiden NMI 0.8252; ΔNMI = +0.1627.

CellBench Drop-seq (3-line mixture): MMTB 3-T local-neighborhood readout (Hamming kNN) 96.89% ± 2.27% (5-fold CV); best MMTB clustering NMI 0.8596 (500 B). Leiden NMI 0.8372; ΔNMI = +0.0224.

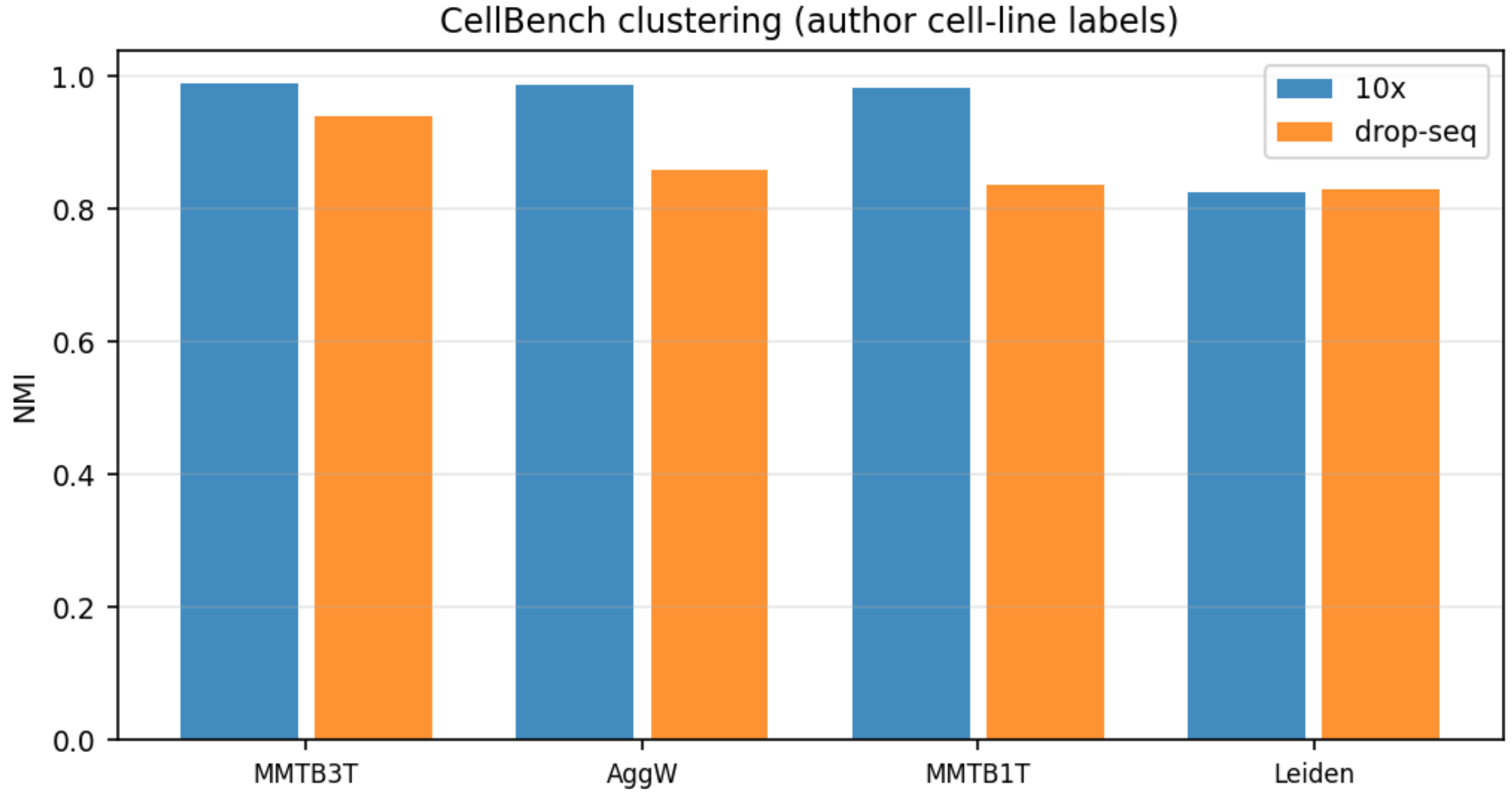

**Figure 2.** Category I clustering agreement on CellBench (author annotations). Bar plot comparing MMTB 3-T Ward-on-bits NMI against Leiden PCA30 NMI for CellBench 10x (NMI: 0.988 vs 0.825) and CellBench Drop-seq (NMI: 0.860 vs 0.837). MMTB 3-T exceeds Leiden on the 10x platform and is competitive on Drop-seq.

Results evaluate representation adequacy under each method's native feature space.

**Interpretation.** CellBench is a low-complexity mixture of three distinct cancer cell lines and represents an upper-bound favorable scenario where thermometer fingerprints align with strong, sparse differential signals. **Table 9** (CellBench row) reports Ward-on-bits NMI = 0.9879 as a hierarchical readout of the same 3-T code; it should be read together with the fair graph comparison in **Table 8** and should not be extrapolated to all sparse-count domains.

### 4.1.3 Challenging biological structure: MicroPheno body-site labels

MicroPheno (HMP 6-mer body sites, 1,192 samples, author body-site labels) is a harder cross-domain setting than the CellBench mixture. On this cohort MMTB 1-T (NMI ≈ 0.79) is preferable to 3-T; label-free tier routing selects 1-T for dense 6-mer rates. Leiden or PCA often remains competitive or ahead; reported as a limitation of the thermometer preset under body-site structure, not as co-primary evidence (**Table 9**; Supplementary Note 6).

MicroPheno (author body-site labels): see **Table 9** and Supplementary Note 6 for complete rankings. The dataset has suitability score $S = 80$, $\bar{E} = 0.2673$, MMTB 3-T NMI = 0.6409, Leiden NMI = 0.6731, and the best alternative (Louvain PCA30) reaches 0.7124, giving ΔNMI = −0.0322 for the 3-T preset versus Leiden.

**Interpretation.** MicroPheno requires distinguishing five human body sites from 6-mer k-mer profiles. MMTB 1-T (NMI ≈ 0.79) outperforms 3-T in this dense k-mer setting, demonstrating the importance of tier routing. That Leiden (0.673) nonetheless exceeds MMTB 3-T (0.641) and competes with MMTB 1-T suggests thermometer presets are not universally optimal across feature types; k-mer profiles with different sparsity structure may benefit from different threshold calibration or continuous embeddings.

#### 4.1.4 Challenging biological structure: PBMC 3k Seurat marker annotations

For PBMC panels that lack accepted author cell-type labels in our Ward-reference protocol, we re-score PBMC 3k against independent Seurat marker-based annotations (K=9; **Table 6**) as a limitation exemplar for tissue-like subtype recovery. MMTB 3-T Ward-on-bits NMI = 0.392 versus Scanpy/Seurat-like Leiden ≈ 0.489: continuous graph pipelines lead on this annotation. We report the gap candidly; it does not elevate Category III as co-primary success evidence.

**Interpretation.** PBMC 3k with Seurat marker annotations (K=9) is a more realistic tissue-subtype recovery task than CellBench. The NMI gap (MMTB 0.392 vs Scanpy/Leiden 0.489) illustrates a core limitation: thermometer fingerprints, with their fixed three-threshold ordinal resolution, capture less of the fine-grained expression variation needed to distinguish closely related immune cell types. This gap defines a key boundary of MMTB's applicability: it is better suited to coarse separation problems than to fine subtype discovery.

#### 4.1.5 Binary baseline comparison (Ward-on-bits)

**Table 8** (columns 3–5) compares MMTB 3-T against training-free binary baselines (SimHash-500; Median-bit on MinMax) under the same Ward-on-bits readout on independent cohorts (PBMC 4k development excluded). All binary codes use Ward-on-bits; Median-bit thresholds are per-feature medians of MinMax-normalized values.

On CellBench 10x, thermometer, SimHash, and Median-bit are competitive near ceiling; on MicroPheno, Median-bit and MMTB 1-T can exceed 3-T while Leiden leads overall; on PBMC 3k Ward-ref, MMTB 3-T beats SimHash and Median-bit. PCA30 (120 B) is more compact than packed 3-T (188 B) but continuous; ~10× storage claims are versus dense 500-D float32 (2000 B), not versus PCA30.

Complete results for SimHash and Median-bit baselines under Ward-on-bits are provided in Supplementary Table S21.

### 4.1.6 Neighborhood preservation (supervised Hamming kNN)

The k-NN readout reported here uses the MMTB binary code as a feature representation, but the classification itself is supervised (requires training labels). The goal is to evaluate whether the binary representation preserves local neighborhood structure, not to claim a fully unsupervised labeling capability. This is a standard evaluation protocol for representation learning (see Section 3.5). The same binary code supports local-neighborhood readout via Hamming kNN (Eq. 7); we verify that the binary representation preserves cell-type neighborhoods on Category I labels, not that it beats trained float classifiers. **Table 10** lists MMTB 1-T and 3-T presets only (5-fold stratified CV; fold-specific Min-Max). On CellBench author annotations, both presets exceed 96%.

**Table 10.** Local neighborhood preservation: Hamming kNN on MMTB binary code (5-fold CV; MMTB presets only)

| Dataset | Preset | Mean Acc (%) ± SD | Storage |
|---|---|---|---|
| CellBench 10x | 3-T | 99.67 ± 0.44 | 188 B |
| CellBench 10x | 1-T | 99.67 ± 0.44 | 500 B |
| CellBench Drop-seq | 3-T | 96.89 ± 2.27 | 188 B |
| CellBench Drop-seq | 1-T | 97.33 ± 2.59 | 500 B |
| PBMC 4k | 3-T | 93.23 ± 0.56 | 188 B |
| PBMC 4k | 1-T | 94.11 ± 0.18 | 500 B |

Because the objective of this experiment is to evaluate local neighborhood preservation of the binary representation rather than supervised classifier optimization, only Hamming k-NN is reported.

## 4.2 Consistency with reference partitions (Secondary analysis)

Consistency checks against Ward-derived reference partitions (Category II: PBMC panels, Moignard, Paul 2015) evaluate representational agreement under a standardized hierarchical objective. Because MMTB's primary clustering readout also uses agglomerative linkage related to Ward, these checks are partly "apples-to-apples" by design and are not independent biological validation. **Figure 3** visualizes the NMI ranking of MMTB variants against PCA-based baselines on PBMC 4k under the Ward reference partition. MMTB 3-T (NMI = 0.8177) and 1-T (NMI = 0.8550) are competitive with Spectral (0.8664) and Agg Ward PCA (0.8317). This is a consistency diagnostic, not biological validation; the reference partition generator (Agg Ward Log1p 500D, NMI = 1.000) is excluded from the plot for fair comparison.

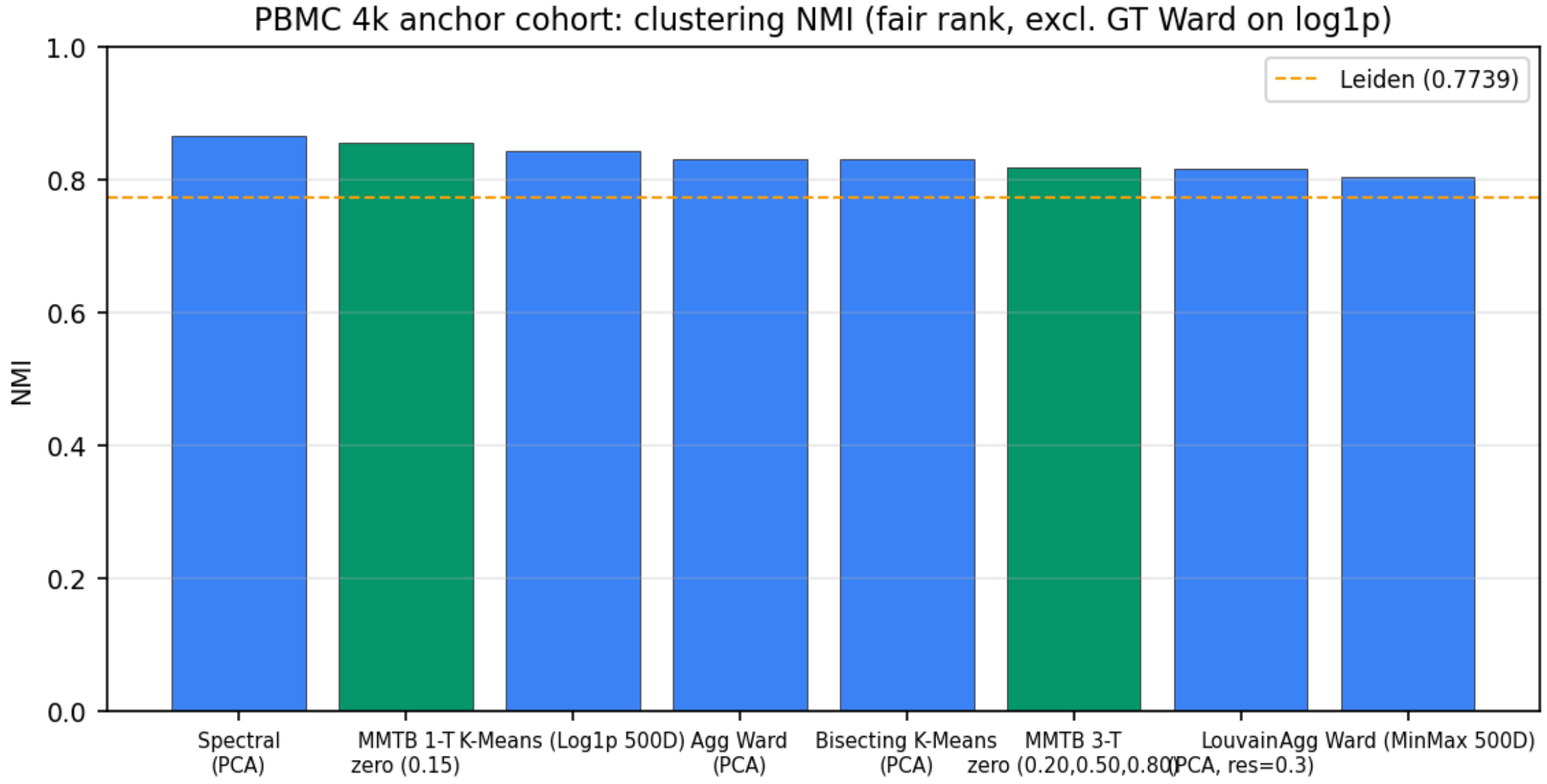

**Figure 3.** PBMC 4k clustering NMI ranking against Ward reference partition (Category II). Bar plot comparing NMI scores of MMTB variants and baseline algorithms on PBMC 4k under a standardized Ward reference partition ($K = 5$). MMTB 3-T (NMI = 0.8177) and MMTB 1-T (NMI = 0.8550) are competitive with PCA-based baselines including Spectral (0.8664) and Agg Ward PCA (0.8317), while falling below the reference generator Agg Ward Log1p 500D (NMI = 1.000, excluded from fair ranking). Category II is a consistency check, not biological ground truth. See Supplementary Note 6 for full rankings and Table S15.

Summary and full rankings, including the former main-text Table 10 assets, are reported in Supplementary Note 6 (Tables S15–S16 and Table S19 summary). Primary biological claims rest on Category I; the independent PBMC check is Category III (Seurat marker annotations, Section 4.1.4). Category II therefore demonstrates self-consistency under shared hierarchical objectives rather than external biological validity.

## 4.3 Threshold Selection and Capacity Trade-off

Preset thresholds provide a stable capacity-agreement trade-off under non-learned feature extraction constraints (Supplementary Note 1). The 3-T preset offers the best balance between ordinal capacity, robustness, and downstream consistency across datasets, and is therefore adopted as the default configuration. Beyond 3-T, capacity continues to grow but downstream NMI gains are marginal and come at the cost of larger fingerprints and greater sensitivity to threshold placement.

## 4.4 Robustness to Perturbations

We assess stability under synthetic dropout, feature dropout, and count-realistic Poisson and mild Negative-Binomial UMI resampling (log1p(Poisson(scale·expm1(X))) from the log1p HVG matrix; Poisson and Negative-Binomial UMI resampling additive noise is no longer used). Batch-effect transfer is left as future work. Summary margins appear in Table 11; SimHash and Median margins under the same perturbations are in the Supplementary Information.

**Table 11.** Robustness on PBMC 3k subset (MMTB 3-T vs Leiden)

| Perturbation | MMTB 3-T NMI | Leiden NMI | Δ NMI |
|---|---|---|---|
| clean | 0.7743 | 0.7295 | +0.0448 |
| dropout_10pct | 0.6963 | 0.7088 | -0.0125 |
| dropout_30pct | 0.5055 | 0.5549 | -0.0494 |
| dropout_50pct | 0.2791 | 0.3572 | -0.0781 |
| poisson_scale_0.5 | 0.7717 | 0.7235 | +0.0482 |
| poisson_scale_1.0 | 0.7195 | 0.7532 | -0.0336 |
| poisson_scale_2.0 | 0.7640 | 0.7524 | +0.0116 |
| nb_theta_5 | 0.7025 | 0.7276 | -0.0251 |
| nb_theta_20 | 0.7340 | 0.7419 | -0.0079 |
| feature_dropout_10pct | 0.7652 | 0.7613 | +0.0039 |
| feature_dropout_30pct | 0.7206 | 0.7142 | +0.0064 |

Baseline MMTB NMI = 0.7743, Leiden = 0.7295. Worst MMTB drop vs clean: -0.4952; min MMTB−Leiden margin: -0.0781.

## 4.5 Supplementary validation and negative controls

Additional analyses support the stability, geometric correspondence, and domain boundaries of the MMTB representation. Bootstrap resampling confirms stable Category II consistency under repeated subsampling (Supplementary Note 2). Empirical Hamming-L1 correlations support the geometric correspondence between thermometer Hamming distance and normalized L1 (Supplementary Note 3). Negative controls on non-count and low-discrimination data, including 20 Newsgroups and RCV1, delineate domains outside the intended operating regime (Supplementary Note 4). These supplementary analyses collectively reinforce the primary representation-level evaluations in Section 4.1 without altering the main conclusions. Bootstrap resampling indicates stable Category II consistency under repeated subsampling (Supplementary Note 2).

## 4.6 Suitability Screening

Label-free suitability screening, comprising score $S$, threshold discrimination $\bar{E}$, and 1-T versus 3-T tier routing, serves as a practical guideline for assessing whether a given sparse count matrix is likely to benefit from MMTB before any labels are used (**Figures 4–5**). $S$ is an engineering heuristic designed for conservative deployment rather than a proven statistical predictor.

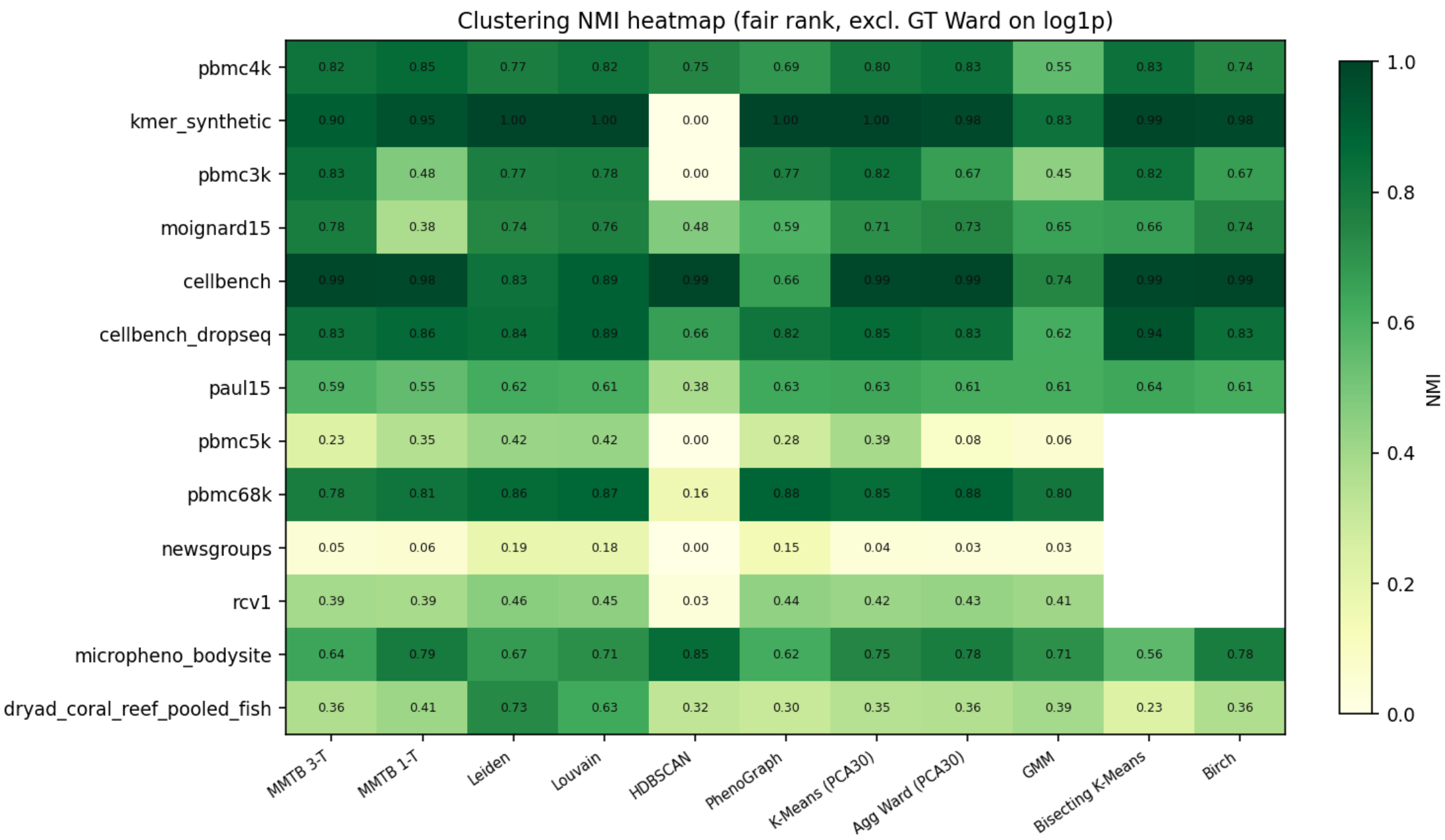

**Figure 4.** Cross-dataset clustering heatmap for suitability-conditioned evaluation. Rows represent datasets; columns represent algorithms. Color intensity indicates NMI.

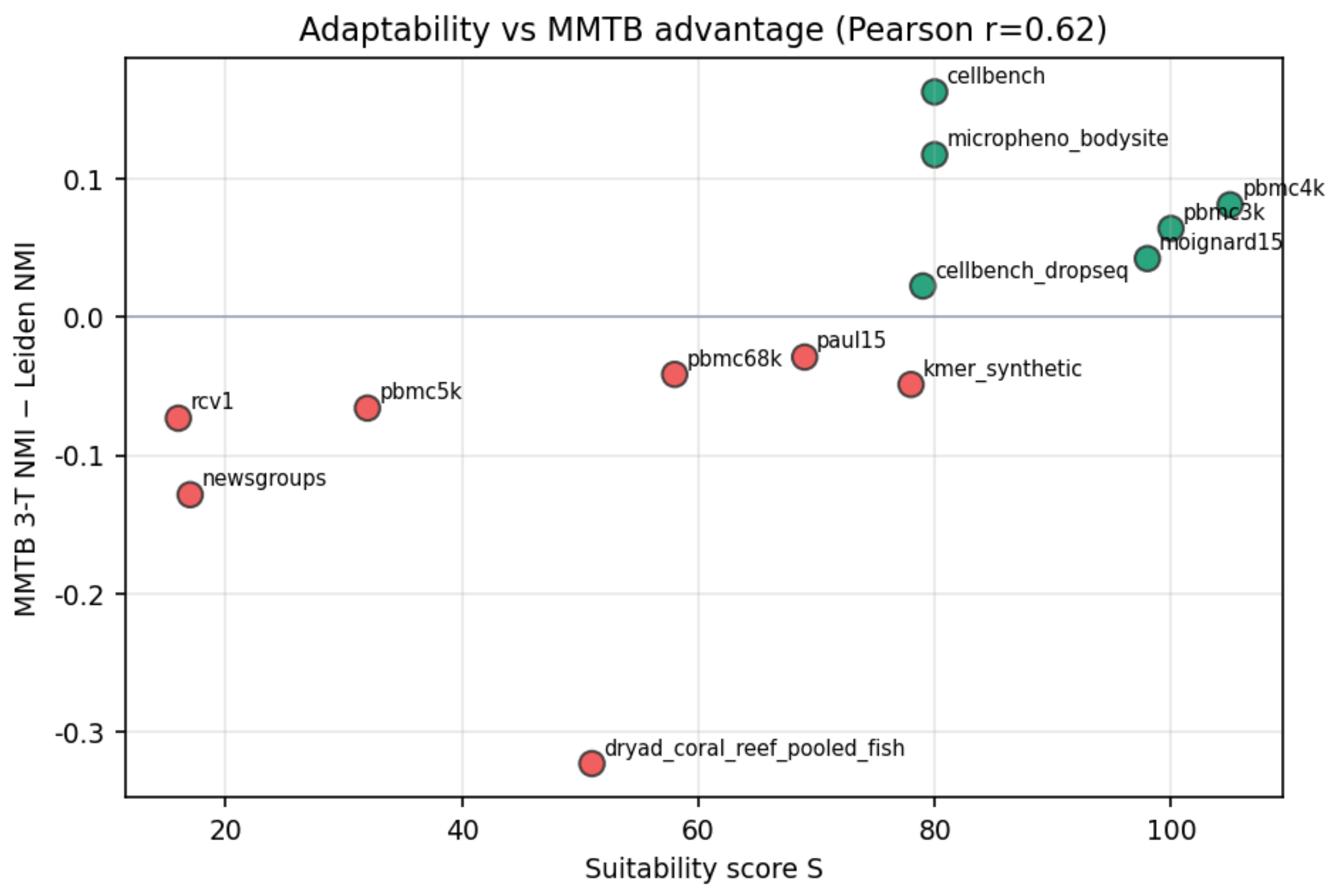

**Figure 5.** Suitability score versus MMTB–Leiden ΔNMI (label-free screening). Each point corresponds to a dataset. The dashed line indicates a descriptive trend; $S$ is a heuristic guideline, not a cross-validated predictor.

Across the 12 evaluation cohorts, the descriptive Pearson correlation between $S$ and ΔNMI is $r = 0.619$. Leave-one-dataset-out recomputation of the same correlation on each $n-1$ subset yields mean $r = 0.6265 \pm 0.0443$ (range 0.578–0.7612). These summary statistics support $S$ as a screening heuristic ("worth testing") rather than a formally validated predictor; weights were not optimized by cross-validation, and no performance guarantee is implied for any particular matrix. In practice, users should treat $S \geq 65$ and $\bar{E} \geq 0.20$ as eligibility indicators for testing MMTB, not as guarantees of downstream performance.

## 4.7 Cross-Domain Sparse Count Benchmarks

Additional sparse-count domains from ecology and natural language processing are summarized in **Table 12**. MicroPheno (microbiome k-mer) results are reported in Section 4.1.3; complete rankings for all cross-domain datasets appear in Supplementary Note 6.

**Table 12.** Cross-domain clustering (MMTB 3-T vs Leiden)

| Dataset | Domain | $S$ | $\bar{E}$ | MMTB 3-T NMI | Leiden NMI |
|---|---|---|---|---|---|
| Synthetic k-mer counts (n=2000) | k-mer mixture | 78 | 0.2110 | 0.8991 | 1.0000 |
| 20 Newsgroups (BoW, n=2000) | NLP text | 17 | 0.0446 | 0.0478 | 0.1856 |
| RCV1 Reuters (BoW, n=2000) | NLP text | 16 | 0.0591 | 0.3911 | 0.4599 |
| Dryad pooled reef fish counts (120 sites, 5 regions) | ecology | 51 | 0.1693 | 0.3645 | 0.7336 |
| Synthetic k-mer counts (n=2000) | k-mer mixture | 78 | 0.2110 | 0.8991 | 1.0000 |

These results are consistent with the suitability screening heuristic: datasets with $S < 65$ or $\bar{E} < 0.20$ (Dryad, 20 Newsgroups, RCV1) show substantial performance gaps relative to graph-based continuous baselines, while the synthetic k-mer dataset ($S = 78$) shows competitive MMTB performance, though Leiden still leads. This pattern reinforces that MMTB is best suited to sparse count matrices with adequate threshold discrimination, and is not recommended for domains with collapsed or extremely sparse binary activation patterns.

## 4.8 Runtime, Memory, and Scaling

Fingerprint generation is $O(Nd)$; end-to-end clustering with Ward linkage is $O(N^2dm)$ and becomes the bottleneck for $N >\sim 10{,}000$. Runtime measurements in Supplementary Note 5 separate encode-only from encode-plus-clustering time; these represent different computational stages and should not be interpreted as direct end-to-end speedups against a fully matched continuous pipeline. Fingerprint generation scales roughly linearly with sample size; end-to-end clustering time depends on the chosen algorithm (Supplementary Note 5 / Tables S9–S12).

# 5. Discussion and Limitations

Practical objective. MMTB is a resource-efficient representation layer for sparse count matrices, intended for scenarios where deterministic inference, compact storage, and non-learned feature extraction are more important than maximizing clustering accuracy under continuous embeddings. It is a complementary tool for constrained settings, not a replacement for PCA or graph-based pipelines when maximal accuracy and abundant compute are available.

Scope of validation. CellBench illustrates a favorable low-K mixture where binary fingerprints can match continuous baselines; MicroPheno and PBMC Seurat markers show settings where continuous pipelines often lead. Category II evaluates consistency with standardized reference partitions only. Suitability score $\boldsymbol{S}$ is a user-guide screen (worth testing), not performance evidence. Highly Suitable matrices ($\boldsymbol{S} \geq \mathbf{80}$) flag deployment candidates for testing; remaining cohorts are boundary or negative controls.

## 5.1 Why Multi-Threshold Works

Single-threshold binarization caps per-feature information at 1 bit (Eq. 13) and collapses when $\boldsymbol{p_j} \to \mathbf{0}$ or $\mathbf{1}$. Thermometer encoding with $\boldsymbol{m}$ nested cutoffs raises the ordinal capacity to $\mathbf{log_2}(\boldsymbol{m}+\mathbf{1})$ bits per feature (Eq. 11) and increases the mutual-information upper bound $\boldsymbol{I}(\boldsymbol{L}; \boldsymbol{C})$ (Eq. 14). Hamming distance on nested bits counts threshold crossings per feature (Eq. 8), yielding a multi-resolution summary of normalized L1 separation (Eqs. 9–10). Threshold ablation and sensitivity analyses on the PBMC 4k reference-clustering panel support the fixed presets $\{\mathbf{0.20}, \mathbf{0.50}, \mathbf{0.80}\}$ (Supplementary Note 1).

Default threshold count. Although 2-T already achieves competitive clustering accuracy (Table 15 in Supplementary Note 1), the 3-T preset offers the best balance between ordinal capacity, robustness, and downstream consistency across datasets, and is therefore adopted as the default configuration. Beyond 3-T, capacity continues to grow but downstream NMI gains are marginal and come at the cost of larger fingerprints and greater sensitivity to threshold placement.

## 5.2 When to Use 1-T vs 3-T

Preset routing (suitability.py) is a heuristic map from matrix traits to 1-T versus 3-T (**Table 13**), not a hard assurance. CellBench 10x favors 3-T; Drop-seq (small cells-per-cluster) can favor 1-T.

**Table 13.** Label-free 1-T vs 3-T routing (zero-training feature-extraction presets)

| Scenario | Condition | Clustering | Neighborhood readout |
|---|---|---|---|
| Coarse scRNA-seq | $K \leq 10$ and cpc $\geq 200$ | 3-T | 3-T |
| Small sample / noisy | cpc $< 100$ | compare | compare |
| Fine subtypes | $K > 10$ | 3-T | 1-T |
| weak_T23 | cpc>=400 AND E2<0.22 AND E31-T1-T | 1-T | 1-T |
| low_E | E<0.20 OR S_td<=8 | 1-T | 1-T |

Routing rationale. Rules are heuristic. Numeric basis: CellBench 10x (author labels, $K = 3$, cpc $\approx 301$) — 3-T clustering NMI 0.9879; CellBench Drop-seq ($K = 3$, cpc $= 75$) — 1-T neighborhood readout can match or exceed 3-T. Borderline matrices ($100 \leq$ cpc $< 200$) should compare both presets. PBMC 4k ablation under a Ward reference is informative for routing, not primary biological validity.

## 5.3 Limitations

Zero-training refers to fingerprint construction only. Ward or agglomerative clustering and Hamming k-NN are downstream consumers whose wall-clock and $O(N^2)$ memory cost dominate at large $N$; MMTB is best viewed as a fixed, compact feature extractor for Hamming geometry, not a free end-to-end clusterer.

**Boundary case:** Paul 2015. Paul 2015 (mouse bone marrow, $K = 19$) passes screening gates ($S \approx 69$, $\bar{E} \approx 0.38$) yet MMTB underperforms Leiden (NMI 0.5917 versus 0.6207; $\Delta$NMI $\approx -0.029$). High $K$ interacting with thermometer capacity is a concrete limitation: eligibility or "worth testing" does not guarantee outperformance.

**Several limitations should be stated explicitly.** First, Ward-reference evaluations (PBMC, Moignard, Paul, PBMC 68k) are consistency checks rather than biological ground truth. Second, the suitability score $S$ is an engineering heuristic rather than an optimal statistical estimator; weights prioritize conservative deployment and $S$ only suggests, does not prove, applicability. Third, fixed threshold presets may not be optimal for every sparse-count domain. Fourth, graph-based continuous embeddings remain advantageous for some matrices, especially fine subtypes, strong batch structure, or low $\bar{E}$.

Fingerprint generation is $O(Nd)$; end-to-end clustering with Ward linkage is $O(N^2dm)$ and becomes the bottleneck for $N >\sim 10{,}000$. Speedups quoted in this work separate encode-only from encode-plus-clustering time; see Supplementary Note 5 / Table S12.

Preset thresholds $\{0.20, 0.50, 0.80\}$ (3-T) and $\{0.15\}$ (1-T) are engineering heuristics selected on the PBMC 4k development panel to maximize label-free $\bar{E}$ (Section 2.4); they are not claimed to be optimal across domains. PBMC 4k is development data for threshold choice; Category II numbers for this panel appear only in Supplementary Note 6. Feature distributions that diverge from the development setting can change the preferred tier; for example, on MicroPheno (author body-site labels), MMTB 1-T (NMI 0.790) is preferable to 3-T. Before deploying MMTB, users should run the label-free screen (score $S$ and threshold discrimination $\bar{E}$). Treat $S \geq 65$ and $\bar{E} \geq 0.20$ as "worth testing," not as a performance guarantee. When $S$ or $\bar{E}$ is low, prefer 1-T or retain a continuous embedding pipeline.

**Clarification of storage comparisons.** Throughout this work, "approximately 10-fold reduction" refers to compressing the original 500-dimensional float32 log1p matrix (2000 B/cell) to the MMTB 3-T packed binary fingerprint (188 B/cell). This is the appropriate baseline because both are full-feature representations ($d = 500$). PCA30 (120 B/cell) is a different representation type, a trained low-dimensional projection. We do not compare binary fingerprints directly to PCA embeddings in our efficiency claims, as such comparisons conflate representation learning strategy with compression.

**Resource profile.** On suitable cohorts the design favors compact deterministic fingerprints and non-learned representation construction relative to the original 500-dimensional float matrix. Encoding and fingerprint-generation cost should not be equated with end-to-end clustering cost, which depends on the selected clustering algorithm (Supplementary Note 5). For $N \lesssim 5{,}000$, Ward linkage on binary fingerprints is practical; for $N \gtrsim 10{,}000$, Hamming kNN graph construction or alternative scalable clustering should be used.

NMI versus storage and non-learned feature extraction. Downstream NMI is a readout of representational adequacy, not a mandate to pick the highest-scoring clustering algorithm. On CellBench 10x (author labels), Agg Ward (PCA30) reaches NMI = 0.993 versus MMTB 3-T 0.988, but requires PCA fitting and float32 embeddings; MMTB 3-T uses 188 B packed bits with no training step. The design trade-off is fixed, lightweight fingerprints that preserve sufficient neighborhood structure, not maximal NMI under every objective and metric.

Positioning. MMTB should therefore be viewed as a resource-efficient complementary tool for sparse count matrices, with demonstrated utility on suitable cohorts. For applications where maximal clustering accuracy is the sole objective and computational resources are abundant, continuous embedding pipelines remain the standard approach.

A recurring pattern in our results is that MMTB performs best on coarse, low-complexity separation tasks (CellBench) and less well on fine subtype recovery (PBMC Seurat markers). This is not coincidental: thermometer thresholds collapse each feature to at most $m + 1$ ordinal levels ($m = 3$ for 3-T, so four levels). This resolution is sufficient to separate cell types with strong, sparse marker differences (cancer cell lines), but may be inadequate for closely related immune subtypes where expression differences are continuous and subtle. This inherent capacity-resolution trade-off (Section 2.4) is a deliberate design choice for compactness and speed, and defines the intended use case rather than a flaw per se.

### 5.4 Future Work

Future work includes: (1) GPU-accelerated Ward or Hamming graph construction for $N > 10^4$; (2) per-domain preset calibration without label leakage; (3) integration with spatial and multi-omic count matrices. Beyond clustering and kNN labeling, MMTB can serve as a lightweight indexing layer for large-scale approximate search and memory-efficient retrieval of sparse count matrices; Hamming distance on packed fingerprints enables fast candidate filtering before expensive float similarity or graph construction.

## 6. Conclusion

MMTB is not a replacement for PCA or graph-based biological clustering. It is a deterministic, compact, non-learned binary representation layer that enables Hamming-indexable analysis of sparse count matrices under resource constraints.

Representative performance is mixed by design. Under a shared Leiden readout, thermometer Hamming graphs can approach PCA+Leiden on favorable coarse mixtures (CellBench) while trailing on tissue-marker labels (PBMC Seurat). Hierarchical Ward-on-bits illustrates an upper-bound favorable scenario and is reported alongside, not instead of, the fair graph readout.

Compared with dense float32 representations, compact fingerprints reduce memory requirements by approximately 10-fold while providing fixed-width Hamming-indexable codes. PCA30 embeddings (120 B/cell) and sparse CSR of the original counts may be smaller; we do not claim universal compression. Suitability score $S$ (Section 3.1) is a deployment guideline only, not a performance predictor.

The recurring pattern that MMTB excels at coarse separation but underperforms on fine subtype recovery reflects a deliberate capacity-resolution trade-off: thermometer thresholds collapse each feature to at most $m + 1$ ordinal levels, sufficient for strong sparse signals but limited for subtle continuous variation. This defines the intended use case rather than a flaw.

Future work includes independent validation of $S$ on external collections, per-domain threshold calibration without label leakage, and GPU-accelerated Hamming graph construction for $N > 10^4$. Beyond clustering and kNN labeling, MMTB can serve as a lightweight indexing layer for large-scale approximate search and memory-efficient retrieval of sparse count matrices.

**Code and Data Availability.**

Source code is available at: https://github.com/tyrone1979/MMTB.

# References

Haque, A., Engel, J., Teichmann, S. A., & Lonnberg, T. (2017). A practical guide to single-cell RNA sequencing for biomedical research and medicine. Genome Medicine , 9(1), 75. https://doi.org/10.1186/s13073-017-0467-8

Pedregosa, F., Varoquaux, G., Gramfort, A., et al. (2011). Scikit-learn: Machine learning in Python. Journal of Machine Learning Research , 12, 2825-2830. https://jmlr.org/papers/v12/pedregosa11a.html

Ng, A. Y., Jordan, M. I., & Weiss, Y. (2001). On spectral clustering: Analysis and an algorithm. Advances in Neural Information Processing Systems , 14, 849-856. https://papers.nips.cc/paper/2092-on-spectral-clustering-analysis-and-an-algorithm

Blondel, V. D., Guillaume, J. L., Lambiotte, R., & Lefebvre, E. (2008). Fast unfolding of communities in large networks. Journal of Statistical Mechanics: Theory and Experiment , 2008(10), P10008. https://doi.org/10.1088/1742-5468/2008/10/P10008

Traag, V. A., Waltman, L., & van Eck, N. J. (2019). From Louvain to Leiden: guaranteeing well-connected communities. Scientific Reports , 9(1), 5233. https://doi.org/10.1038/s41598-019-41695-z

Stuart, T., Butler, A., Hoffman, P., et al. (2019). Comprehensive integration of single-cell data. Cell , 177(7), 1888-1902. https://doi.org/10.1016/j.cell.2019.05.031

Wolf, F. A., Angerer, P., & Theis, F. J. (2018). SCANPY: large-scale single-cell gene expression data analysis. Genome Biology , 19(1), 15. https://doi.org/10.1186/s13059-017-1382-0

Gong, Y., Lazebnik, S., Gordo, A., & Perronnin, F. (2013). Iterative quantization: A procrustean approach to learning binary codes for large-scale image retrieval. IEEE Transactions on Pattern Analysis and Machine Intelligence , 35(12), 2916-2929. https://doi.org/10.1109/TPAMI.2012.193

Shen, F., Shen, C., Liu, W., & Tao Shen, H. (2015). Supervised discrete hashing. Proceedings of the IEEE Conference on Computer Vision and Pattern Recognition , 37-45. https://doi.org/10.1109/CVPR.2015.7298688

Zheng, G. X., Terry, J. M., Belgrader, P., et al. (2017). Massively parallel digital transcriptional profiling of single cells. Nature Communications , 8, 14049. https://doi.org/10.1038/ncomms14049

Xu, C., & Su, Z. (2024). scCluBench: a comprehensive benchmarking framework for clustering algorithms on single-cell RNA sequencing data. bioRxiv , 2024.03.15.585042. https://doi.org/10.1101/2024.03.15.585042

Tian, L., Su, S., Amann-Zalcenstein, D., et al. (2019). scRNA-seq mixology: towards a robust sample multiplexing schema for single-cell RNA sequencing studies. Genome Biology , 20, 201. https://doi.org/10.1186/s13059-019-1882-2

Breiman, L. (2001). Random forests. Machine Learning , 45(1), 5–32. https://doi.org/10.1023/A:1010933404324

Cover, T. M., & Hart, P. E. (1967). Nearest neighbor pattern classification. IEEE Transactions on Information Theory , 13(1), 21–27. https://doi.org/10.1109/TIT.1967.1053944

Cortes, C., & Vapnik, V. (1995). Support-vector networks. Machine Learning , 20(3), 273–297. https://doi.org/10.1007/BF00994018

Ward, J. H. (1963). Hierarchical grouping to optimize an objective function. Journal of the American Statistical Association , 58(301), 236–244. https://doi.org/10.1080/01621459.1963.10500845

Strehl, A., & Ghosh, J. (2002). Cluster ensembles — a knowledge reuse framework for combining multiple partitions. Journal of Machine Learning Research , 3, 583–617 (NMI). https://jmlr.org/papers/v3/strehl02a.html

Hubert, L., & Arabie, P. (1985). Comparing partitions. Journal of Classification , 2(1), 193–218 (ARI). https://doi.org/10.1007/BF01908075

Charikar, M. S. (2002). Similarity estimation techniques from rounding algorithms. Proceedings of the 34th ACM Symposium on Theory of Computing , 380–388 (SimHash). https://doi.org/10.1145/509907.509965

Gionis, A., Indyk, P., & Motwani, R. (1999). Similarity search in high dimensions via hashing. Proceedings of the 25th VLDB Conference , 518–529 (LSH). https://doi.org/10.1016/S0079-9994(08)60152-2

Hubara, I., Courbariaux, M., Soudry, D., El-Yaniv, R., & Bengio, Y. (2016). Quantized neural networks: Training neural networks with low precision weights and activations. Journal of Machine Learning Research , 18(187), 1–30 (thermometer encoding). https://doi.org/10.5555/3122009.3242044

Jégou, H., Douze, M., & Schmid, C. (2011). Product quantization for nearest neighbor search. IEEE Transactions on Pattern Analysis and Machine Intelligence , 33(1), 117–128. https://doi.org/10.1109/TPAMI.2010.57

Asgari, E., & Mofrad, M. R. K. (2018). MicroPheno: predicting environments and host phenotypes from 16S rRNA gene sequencing using k-mers and machine learning. Bioinformatics , 34(13), i32–i42. https://doi.org/10.1093/bioinformatics/bty296

Connolly, S. R., MacNeil, M. A., & Caley, M. J. (2017). Real-world complexity versus meta-community model dynamics. Ecology Letters , 20(4), 494–504. https://doi.org/10.5061/dryad.2qp80

McMurdie, P. J., & Holmes, S. (2013). phyloseq: An R package for reproducible interactive analysis and graphics of microbiome census data. PLOS ONE , 8(4), e61217. https://doi.org/10.1371/journal.pone.0061217